\documentclass{article}
\usepackage{ijcai18}

\usepackage{times}
\usepackage{xcolor}
\usepackage{soul}
\usepackage[utf8]{inputenc}
\usepackage[small]{caption}

\usepackage{url}
\usepackage{algorithmic}
\usepackage{algorithm}
\usepackage{graphicx}
\usepackage{amssymb}
\usepackage{color}
\usepackage{enumitem}
\usepackage{amsmath}

\newcommand{\citet}[1]{\citeauthor{#1}~\shortcite{#1}}
\newcommand{\citep}{\cite}

\newcommand{\method}{{\sc CHAD}}

\title{Detecting Changed-Hands Online Review Accounts}

\iftrue
\author{
Geli Fei\dag, 
Shuai Wang\dag, 
Bing Liu\dag,
Leman Akoglu\ddag 
\\ 
\dag Department of Computer Science,
University of Illinois at Chicago, USA \\
\ddag Heinz College of Information Systems and Public Policy,
Carnegie Mellon University, USA  \\
\{gfei2,swang207,liub\}@uic.edu,
lakoglu@andrew.cmu.edu
}
\fi

\begin{document}

\maketitle

\begin{abstract}
  A reputable social media or review account can be a good cover for spamming activities. It has become prevalent that spammers buy/sell such accounts openly on the Web. We call these sold/bought accounts the \textit{changed-hands (CH) accounts}. They are hard to detect by existing spam detection algorithms as their spamming activities are under the disguise of clean histories. {In this paper, we first propose the problem of detecting CH accounts, and then design an effective detection algorithm which exploits changes in content and writing styles of individual accounts, and a proposed novel feature selection method that works at a fine-grained level within each individual account.} The proposed method not only determines if an account has changed hands, but also pinpoints the change point. Experimental results with online review accounts demonstrate the high effectiveness of our approach.
\end{abstract}

\section{Introduction}

\textit{Opinion spam} has become a common type of spam in review sites such as Amazon and Yelp, as people continue to heavily rely on online reviews to make purchase decisions.
Since the early work by \citet{jindal2008opinion}, detecting fake reviews and reviewers have drawn wide attention from both the research community and the industry. 
The problem has been investigated through different approaches, including those based on content or linguistic information \cite{ott2011finding,li2014towards}, reviewer behaviors \cite{feng2012distributional,ye2015discovering}, temporal posting patterns \cite{xie2012review,kc2016temporal}, and relational analysis \cite{conf/kdd/JiangCBFY14,conf/kdd/RayanaA15}.

As a result of the advances in spam filtering techniques, spamming has become harder than before. For example, giving all-extreme ratings or posting many reviews in a short time frame can be easily caught. Driven by profits, opinion spammers resort to other strategies. One strategy is to offer to buy reputable accounts\footnote{\url{https://www.yelp.com/topic/boston-someone-offered-to-buy-my-yelp-account}} (those with a clean history) and use them to post spam reviews. Selling/buying accounts is also prevalent in other forms of social media. Karma farmers\footnote{\url{https://www.reddit.com/r/AgainstKarmaWhores/comments/383qsp/why\_are\_we\_doing\_this/}} 
are such an example in the community website Reddit, who try to gain high karma (upvotes and reputation) quickly with new accounts so that their posts can show up in the front page, and then sell these seemingly 
reputable accounts to spammers.

In both of the above situations, accounts change hands at a certain time point and they unavoidably exhibit linguistic and writing style 
differences in the midst of their life span. It is hard for a spammer to align his writing style with the original account holder's writing style for two reasons. First, there is no simple manual way to quantify another user's 
writing style in every aspect. Second, since spammers have different objectives than legitimate users, e.g., promoting some products, their 
writing styles change naturally. To the best of our knowledge, such changes have not been studied before. This paper represents the first work on the topic. 

In this paper, we propose this new problem of detecting {\em changed-hands} 
(CH) accounts from a content and writing style perspective. An algorithm, called 
{\method} ({\em CH Accounts Detection}), is proposed to identify if an account has changed hands and to estimate the time point of change if so. In case of a change, we assume there is only one change in an account's life time because once there is a change, it should be detected before a second change happens. {Existing spammer detection methods are not suitable for detecting such accounts, and cannot identify the change point for two reasons. (1) They assume there is a single user behind each account. (2) They examine the overall behavior of each account. For CH accounts, their spamming activities may not be obvious given a clean history. This work thus complements the existing review spam detection settings and algorithms.}

\textbf{Problem Definition}: Given an account $\boldsymbol{A} = \{r_1, r_2, \dots, r_n\}$ with reviews $r_j$ sorted by their posting dates, \method~determines whether a significant linguistic and/or writing style change has occurred starting from a particular review $r_i$ ($1 < i < n$). The algorithm returns $i$ if yes, and returns \textit{none} otherwise.



The problem has two unique challenges:
\begin{enumerate}[leftmargin=*,noitemsep,nolistsep]
	\item \textbf{Inter-user differences}: Different CH accounts 
	exhibit different changes, because not every pair of users has the same differences in their writings. For example, in some CH accounts, the two users can be distinguished by the average length of the words they use.
	In some other CH accounts, the two users may be distinguished by the average sentence length but not by the average word length, because one uses long sentences while the other uses short ones, but both of them mainly use short words.
	\item \textbf{Intra-user variance}: Every review is unique in some way, which results in 
	a certain amount of difference and 
	variance even when compared with other reviews of the same user. 
	However, such differences do not indicate a real 
	changing of hands between two users. 
\end{enumerate}

\noindent
Given these two challenges, a desired detection method needs to perform detection at the account level and adjust itself to different individual accounts. {In this paper, we propose an effective detection algorithm with a novel feature selection method, called~\textit{pivot-level feature selection}, to address these challenges. The key novelty of this feature selection method is that, due to the two challenges, it works at a fine-grained level within each individual account rather than the whole dataset as traditional feature selection methods do.} 

In summary, this paper makes the following contributions:

\begin{enumerate}[leftmargin=*,noitemsep,nolistsep]
	\item 
	It proposes 
	the new problem of detecting CH accounts,
	which have become prevalent in many social media sites, but have not been studied so far. {This new problem complements the existing spammer detection settings}. 
	\item 
	It proposes a novel algorithm, 
	\method, which leverages linguistic evidences and a novel new feature selection algorithm to identify if an account has changed hands during its life time and estimates the change point.
	\item It evaluates \method~on two datasets 
	and show that the proposed approach is highly effective.
\end{enumerate}

\section{Related Work}
\label{sec:related-work}
Our work is related to opinion spam detection, tracking linguistic evolution and change point detection.

\subsection{Opinion Spam Detection}
Since the first work by \citet{jindal2008opinion}, a wide range of techniques have been proposed for detecting spam reviews \cite{li2011learning,li2014spotting,haideceptive}, individual spammers \cite{lim2010detecting,akoglu2013opinion} and spammer groups \cite{mukherjee2012spotting}. However, the use of CH accounts as a new instrument for spamming has not been studied thus far and no techniques are available for their detection.

Among existing techniques, detecting spammer accounts is most relevant to our problem. 
\citet{lim2010detecting} studied users' rating behaviors; \citet{akoglu2013opinion} studied the relational collusion between reviewers and their target products; \citet{mukherjee2013spotting} used a Bayesian approach to modeling the behavioral patterns of spammers and non-spammers. These approaches cannot 
detect CH accounts and pinpoint their change locations
because they examine the overall behavioral patterns of each account. The spamming activities of CH accounts may go undetected given a clean history.

Sockpuppet detection \cite{hosseinia2017detecting} refers to the detection of a single author behind multiple accounts. These methods cannot be directly applied as they regard reviews from one account as written by a single author.

Our work is also related to using linguistic approaches to detecting spamming reviews \cite{ott2011finding,ren2014positive} and loosely related to psycholinguistic deception detection \cite{newman2003lying,perez2015verbal}, as we also use a linguistic-based approach.

\subsection{Linguistic Evolution \& Change Point Detection}
On tracking linguistic evolution across time, \citet{juola2003time} quantified the rate of change in language across two time periods, and \citet{lijffijt2012ceecing} studied lexical stability in a historical corpus. Our work is different because the above works 
compare language from two chosen time periods, while our goal is to estimate the change point from a sequence of documents. Tracking shifts in the meaning of words was studied in \cite{
	mitra2014s,kulkarni2015statistically}. Our work does not study shifts of word meaning but ``shifts in authorship.'' However, authorship attribution and verification methods \cite{koppel2004authorship,sanderson2006short}
cannot be applied as we don't have any training data of the users. 
Change point detection is a core time series analysis problem \cite{taylor2000change}. 
In our work, we adopt the \textit{single change point} detection technique by \citet{chen1999change}, 
as it aligns with our goal of detecting CH accounts.


\section{Proposed CHAD Method}
\label{sec:propose-method}
This section presents the proposed \method~algorithm. 

\subsection{The Overall Algorithm}
\label{lab:main-idea}
{The main idea of \method~is based on the observation that the reviews written by one user are similar among themselves but different from those written by a different user.} The \method~algorithm is outlined in Alg.~\ref{algo:1}, which works on one account at a time. Note that it needs a pre-selected feature set $\boldsymbol{F}$ as input, which is a subset of all features $\boldsymbol{F}^{all}$ (We will explain this shortly). We first introduce the five main steps, and then go into details of each step.
\begin{algorithm} [t]
	\small
	\caption{\textbf{CHAD}}
	\textbf{Input:} Account := $\boldsymbol{A}=\{r_1, r_2, \dots, r_n\}$, \\
	\text{\hspace{30pt} Window size := $K$, Smoothing factor := $\lambda_S$} \\
	\text{\hspace{30pt} Features := $\boldsymbol{F}(\subseteq \boldsymbol{F}^{all}) = \{f_1, f_2, \dots, f_m\}$} \\
	\textbf{Output:} Res := index $i$ (1\textless $i$\textless $n$) or \textit{none}
	\begin{algorithmic}[1]
		\STATE $\boldsymbol{C} := \text{\O}$ $~~$ // $\boldsymbol{C}$ is a multiset for voting.
		\FOR {\textbf{each} $r_i \in \boldsymbol{A}  (1 \le i \le n-K+1)$}
		\STATE $\boldsymbol{S}_i := \text{\O}$ // $\boldsymbol{S}_i$ is a set of similarity sequences for a pivot window.
		\STATE $\textit{pivot-window} := \{r_i, \dots, r_{i+K-1}\}$
		\STATE $\overline{\boldsymbol{A}} :=  \boldsymbol{A} \;{\backslash}\; \textit{pivot-window}$
		\FOR {\textbf{each} $f_j \in \boldsymbol{F}$}
		\STATE $ss_{ij} := \text{compute-sim-seq}(r_i, K, \overline{\boldsymbol{A}}, f_j)$
		\STATE $\boldsymbol{S}_i := \boldsymbol{S}_i \cup \{ss_{ij}\}$
		\ENDFOR
		\STATE $\boldsymbol{S}_i := \text{pivot-level-feature-select}(\boldsymbol{S}_i)$
		\STATE $s_i := \text{aggregate}(\boldsymbol{S}_i)$
		\STATE $c_i := \text{change-point-detect}(s_i)$ // $c_i$ is either a review's temporal index or \textit{none}.
		\STATE $\boldsymbol{C} := \boldsymbol{C} \cup \{c_i\}$ 
		\ENDFOR
		\STATE $\textit{res} := \text{is-change-vote}(\boldsymbol{C})$
		\IF {($\textit{res} \ne \textit{none}$)}
		\STATE $\boldsymbol{C}_{-none} := $ remove \textit{none} elements from $\boldsymbol{C}$
		\STATE $\boldsymbol{C}_{-none}^S := \text{smooth}(\boldsymbol{C}_{-none}, \lambda_S)$
		\STATE $\textit{res} := \text{change-point-vote}(\boldsymbol{C}_{-none}^S)$
		\ENDIF
		\RETURN $\textit{res}$
	\end{algorithmic}
	\label{algo:1}
\end{algorithm}
\setlength{\textfloatsep}{0.1in}

\begin{enumerate}[leftmargin=*,noitemsep,nolistsep]
	\item \textit{Generate similarity sequences} (lines 2-9): {For an input account $\boldsymbol{A}$, this step builds a set of similarity sequences $\boldsymbol{S}_i$ using features in $\boldsymbol{F}$ 
	for a {\em pivot window} of $K$ reviews starting from a review $r_i$.} Each sequence $ss_{ij} \in \boldsymbol{S}_i$ is computed by comparing the similarity of reviews in the pivot window and reviews within a moving window of also size $K$ in the remaining reviews $\overline{\boldsymbol{A}}$ using one ($f_j$) of the features in $\boldsymbol{F}$ (line 7). 
	For a CH account, we expect the similarities to be high when comparing reviews written by the same user, but low across two different users. 
	
	\item \textit{Eliminate noisy features} (line 10) 
	
	\item \textit{Aggregate sequences} (line 11): We aggregate the remaining sequences for each pivot window by averaging the sequences in the resulting $\boldsymbol{S}_i$.
	
	\item \textit{Change-point detection} (line 12): 
	We employ a statistical algorithm for {\em change-point} detection on each aggregated sequence $s_i$ to detect the change point. 
	
	\item \textit{Two-round voting} (line 15-20): 
	We perform two rounds of voting on the change-point detection results on the aggregated sequences of all pivot windows for an account to determine if a changing of hands has occurred and also to identify the final change point.
\end{enumerate}

\noindent
{\bf Global feature pre-selection}: As mentioned before, \method~requires a pre-selected feature set $\boldsymbol{F}$ as input. $\boldsymbol{F}$ is selected globally by running Alg.~\ref{algo:1} without line~10 (pivot-level feature selection) on all accounts of a development set for multiple iterations starting with all features $\boldsymbol{F}^{all}$ as input. Each iteration removes one feature from $\boldsymbol{F}^{all}$ that gives the biggest performance gain in F1 score under the change-point evaluation ($\textit{eval}_\textit{cp}$) (Sec.~\ref{sec:data-and-eval})
. It globally removes those noisy features in $\boldsymbol{F}^{all}$.

\subsection{Features and Similarity Metrics}
\label{lab:seq-gen}

Now we list the set of all features $\boldsymbol{F}^{all}$ used in the \textit{compute-sim-seq} function (line 7) in Alg.~\ref{algo:1}. Features with * produce a single value for reviews in a given window and the rest use the Bag-of-Words (BoW) model. Single value features include {\em average sentence length*} and {\em average token length*}. BoW features include {\em word unigrams}, {\em word bigrams}, {\em Part-of-Speech unigrams}, {\em Part-of-Speech bigrams}, {\em adjectives \& adverbs}, {\em nouns}, {\em function words}, and {\em punctuations}.

To measure the similarity between reviews in two review windows, we use cosine similarity for BoW features. We tried some other measures such as Jaccard similarity but they did not perform well. For single value features, we compute the similarity \textit{sim} using their absolute difference \textit{diff} and normalizing it to [0,1]: 
\begin{equation}
\textit{sim} = 1 / (1 + log(1 + \textit{diff}))
\label{eqn:normalize}
\end{equation}

\begin{algorithm} [t]
	\small
	\caption{\textbf{Pivot-level-Feature-Select}}
	\textbf{Input:} $\boldsymbol{S}_i$ := \{$ss_{i1}$, $ss_{i2}$, \dots, $ss_{iT}$\} \\
	\text{\textbf{Output:} $\boldsymbol{E}$ := the set of selected sequences}
	
	\begin{algorithmic}[1]
		\STATE 	\textit{target} := \text{avg}($\boldsymbol{S}_i$)
		\STATE $ \boldsymbol{E} = \text{\O}$
		\STATE $\boldsymbol{S}_i^{'} = \text{sort}(\{\text{Pc}(ss_{ij} \in \boldsymbol{S}_i, target)\})$ // computes Pearson's correlation (\textit{Pc}) of each $ss_{it} \in \boldsymbol{S}_i$ to \textit{target} and sort in descending order.
		\STATE $\boldsymbol{E} := \boldsymbol{E} \cup \{\boldsymbol{S}_i^{'}[1]\}$ // adds the sequence with highest correlation to $\boldsymbol{E}$.
		\FOR {$j \in \{2 : T\}$}
		\STATE $p :=$ \text{avg}$(\boldsymbol{E})$
		\STATE $l := \boldsymbol{S}_i^{'}[j]$
		\IF {$\text{Pc}(\text{avg}(\boldsymbol{E} \cup \{l\}), \textit{target}) > \text{Pc}(p, \textit{target})$}
		\STATE $\boldsymbol{E} := \boldsymbol{E} \cup \{l\}$
		\ELSE
		\STATE \textbf{break}
		\ENDIF
		\ENDFOR
		\RETURN $\boldsymbol{E}$
	\end{algorithmic}
	\label{algo:2}
\end{algorithm}

\subsection{Pivot-Level Feature Selection}
\label{lab:feature-sel}
Now we describe the \textit{pivot-level-feature-select} function in line 10 of Alg.~\ref{algo:1}. As we pointed out earlier, one of our key challenges is that the writing differences between a pair of users in one CH account may be different from those between other pairs in other CH accounts. Furthermore, each review is unique in some way which can result 
in a certain amount of difference 
when computing similarity with other reviews using some features. Such differences however may not indicate real writing differences between two users. To solve these two problems, we propose to perform pivot-level feature selection (Alg.~\ref{algo:2}). The corresponding similarity sequence of each removed feature is deleted from $\boldsymbol{S}_i$.

The \textit{pivot-level-feature-select} function selects a subset of sequences in $\boldsymbol{S}_i$ through correlation analysis, which is the same as selecting their corresponding features. It first averages all sequences in $\boldsymbol{S}_i$ to construct a \textit{target} sequence (line 1). It then computes Pearson's correlation ($Pc$) of each sequence in $\boldsymbol{S}_i$ with the \textit{target} and sorts the sequences based on correlation strength in descending order (line 3). Line 4 adds the sequence with the highest correlation to the result set $\boldsymbol{E}$. It then goes through the sorted sequence set $\boldsymbol{S}_i^{'}$ and tests if adding another sequence to $\boldsymbol{E}$ would increase $\boldsymbol{E}$'s average's correlation to \textit{target} (lines 5-13). If the correlation improves by the addition, we update $\boldsymbol{E}$; otherwise exit and return $\boldsymbol{E}$.

\begin{figure}[t]
	\centering
	\includegraphics[scale=0.26]{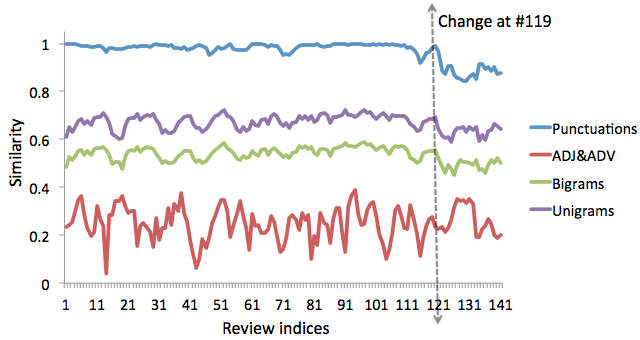}
	\vspace*{-3mm}
	\caption{Some sample sequences in an $\boldsymbol{S}_i$.}
	\label{fig:ss-illustration}
\end{figure}

The intuition is that {\em target} is a representative sequence assuming only a few noisy sequences exist. Through correlation analysis, we identify the sequences that align with the {\em target} and discard those outlying ones that otherwise hinder the change point detection performance. Fig.~\ref{fig:ss-illustration} gives an example of several similarity sequences computed for a pivot window on a CH account in the Amazon dataset
, where the actual change happens at review \#119. For clarity, we only plot a subset of sequences generated using 4 features. Through correlation analysis, our Alg.~\ref{algo:2} is able to effectively eliminate the noisy sequence generated from the feature Adj\&Adv.

\subsection{Change Point Detection}
\label{lab:detecting-change-points}
As mentioned in the introduction, we assume there is at most one change point in each account. As such, we use the single point change detection algorithm by~\citet{chen1999change}, which uses the Schwarz Information Criterion (SIC) to search for the change point. Suppose $X_1, X_2, \dots, X_n$ is a sequence of independent Gaussian random variables with means $\mu_1, \mu_2, \dots, \mu_n$ and variances $\sigma_1^2, \sigma_2^2, \dots, \sigma_n^2$, respectively. The method 
tests the hypothesis of whether there is a single change in both the mean and variance located at the unknown position $k$, $2 \le k \le n-1$ as:

\vspace{-2 mm}
\begin{equation*}
\begin{split}
H_0 : & \mu_1 = \mu_2 = \dots = \mu_n = \mu \text{ and} \\ & \sigma_1^2 = \sigma_2^2 = \dots = \sigma_n^2 = \sigma^2,
\end{split}
\end{equation*}
\noindent versus the alternative hypothesis
\begin{equation*}
\begin{split}
H_1 : & \mu_1 = \dots = \mu_k \ne \mu_{k+1} = \dots = \mu_n \text{ and} \\ & \sigma_1^2 = \dots = \sigma_k^2 \ne \sigma_{k+1}^2 = \dots = \sigma_n^2,
\end{split}
\end{equation*}

\noindent where $\mu$ and $\sigma^2$ are unknown common parameters when there is no change. We thus have two models corresponding to the $H_0$ and $H_1$. The principle of minimum SIC is used to reject $H_0$. In particular, $H_0$ is not rejected if $\text{SIC}(n) \le \text{min}_k \text{SIC}(k)$, and rejected otherwise. $SIC(n')$ is defined 
as $-2 \text{log}L(\hat{\Theta}) + p\text{log}n'$, where $L(\hat{\Theta})$ is the maximum likelihood function for each model, $p$ is the degrees of freedom in the model ($p=2$ under $H_0$ and $p=4$ under $H_1$), and $n'$ is the sample size.

\subsection{Two-Round Voting}
\label{lab:voting}
In Alg.~\ref{algo:1}, \method~uses a two-round voting scheme (step 5) to determine if an account has changed hands and also to pinpoint the location of change (lines 15-20). In the first round (line 15), \textit{is-change-vote} function determines if a change has occurred. Note that each element $c_i \in \boldsymbol{C}$ returned by the change point detection algorithm is either a change point (i.e., a review) or \textit{none} (indicating no change). This function simply counts the number of votes for each change point and \textit{none}. 
If \textit{none} has the highest number of votes, it returns \textit{none}; otherwise it registers that a change has occurred and removes all the \textit{none} elements from $\boldsymbol{C}$ (line 17). It then moves on to the second round of voting to pinpoint the actual change location. Instead of directly voting based on elements in $\boldsymbol{C}_{-none}$, we perform \textit{smoothing} on $\boldsymbol{C}_{-none}$ first (line 18). Let us look at an example. Given a set of votes in $\boldsymbol{C}_{-none}$ in the format of change-point:\#-of-votes 10:8, 50:5, 51:7, 52:4, the point that gets the highest votes is $10$. However, the actual change point is more likely to be around $51$. In order to overcome this possible noise factor, we smooth the votes by adding some extra counts to near-by locations of every change point in $\boldsymbol{C}_{-none}$ to construct $\boldsymbol{C}_{-none}^S$.  Specifically, we pick a smoothing factor $\lambda_S \in \mathbb Z_{>0}$ and for a change point $i$ with $v$ votes, we add $v / \lambda_S^{dist}$ extra votes to locations $i+\textit{dist}$ and $i-\textit{dist}$, where $\textit{dist} = 1,2,\dots$. Finally, we perform the second round of voting on $\boldsymbol{C}_{-none}^S$ to determine the final change location.

\section{Experiments}
\subsection{Datasets and Evaluation Metrics}
\label{sec:data-and-eval}
\textbf{Datasets}: 
For experiments we constructed synthetic datasets for the following reasons: First, no publicly available labeled data exists for our problem; Second, identifying opinion spam manually has been shown to be very unreliable \cite{ott2011finding}; Lastly, although Mechanical Turkers have been used to write individual fake reviews \cite{ott2011finding}, our case is much more complicated because of different sizes and the diversity of reviewed products for different accounts. 
In fact, synthetic data was used before, e.g., in sockpuppet detection \cite{qian2013identifying}. In this work, we use two public review corpora to construct our data, one from Amazon \cite{jindal2008opinion}, which contains reviews for multiple product categories such as books, electronics, etc., and the other from Yelp \cite{mukherjee2013yelp}, which contains only hotel reviews. The construction of CH accounts from each corpus is done as follows: We first randomly select two different original accounts with at least 10 reviews, $\boldsymbol{A}_1=\{r_{11}, r_{12}, \dots, r_{1n}\}$ and $\boldsymbol{A}_2 = \{r_{21}, r_{22}, \dots, r_{2n'}\}$, both sorted by their review posting dates, and then concatenate one account to the other, giving us $\boldsymbol{A}_{syn} = \{r_{11}, r_{12}, \dots, r_{1n}, r_{21}, r_{22}, \dots, r_{2n'}\}$, whose minimum size is 20. We in total constructed 350 CH accounts. Then we 
	sample 350 original accounts with at least 20 reviews as non-CH (NCH) accounts that approximately match the mean and standard deviation of the sizes of the constructed CH accounts by following the 68{-}95{-}99.7 rule from statistics \footnote{\url{https://en.wikipedia.org/wiki/68-95-99.7_rule}}. This way of sampling (rather than random sampling) is important because it ensures that the NCH and CH accounts have similar number of reviews, which eliminates the bias due to joining two accounts in constructing CH accounts that can result in significantly more reviews for CH accounts than for NCH accounts. Thus for each dataset, we in total created 700 accounts. Statistics of the size of the accounts in both datasets are given in Table \ref{table:dataset-stats}. In Sec.~\ref{sec:additional-comparison}
, we will show the results when the datasets are constructed in a different way.

Although it is possible that the original corpora already contain some CH accounts, we believe the chance of selecting existing CH accounts is very small because the original corpora are very large. Also, we believe it is reasonable to study accounts with more than 20 reviews because of our problem setting, i.e., an account changes hand after it has gained enough “reputation” or a long history.

We choose to use the Amazon and Yelp corpora for the following reason. The Amazon corpus has reviews of all kinds of products. We use it to create the scenario where a spammer buys an account and uses it to review products that are potentially very different from those reviewed by the original user (although we do not enforce this when constructing the dataset). Moreover, products reviewed by a single user can also be quite diverse. {In contrast, the Yelp corpus has only hotel reviews, which allows us to show whether our approach can detect CH accounts when the two users wrote reviews for the same type of entities (i.e. when the content change is not as drastic).} As we will show, \method~is able to perform well in both scenarios.

\begin{table}
	\captionsetup{skip=4pt}
	\centering
	\small
	\begin{tabular}{l|c|c|c|c|c}
		\hline
		{} & \textbf{Mean} & \textbf{Med.} & \textbf{Stdev} &\textbf{Min} & \textbf{Max} \\
		\hline
		$Amazon_{NCH}$ & 60.0 & 40 & 37 & 20 & 170 \\
		\hline
		$Amazon_{CH}$ & 53.6 & 41 & 33.2 & 20 & 231 \\
		\hline
		${Yelp}_{NCH}$ & 47.7 & 43 & 14.7 & 20 & 88 \\
		\hline
		${Yelp}_{CH}$ & 45.9 & 41 & 13.7 & 20 & 84 \\
		\hline
	\end{tabular}
	\caption{Review Number Statistics.}
	\label{table:dataset-stats}
\end{table}

\begin{table*}[t]
	\captionsetup{skip=4pt}
	\small
	\centering
	\begin{tabular}{l|c|c|c|c||c|c|c|c}
		\hline
		{} & \multicolumn{4}{c||}{\textbf{$\textit{eval}_\textit{{cha}}$}} & \multicolumn{4}{c}{\textbf{$\textit{eval}_\textit{{cp}}$}} \\
		\hline
		Feature Name & Prec. & Recall & F1 & Accu. & Prec. & Recall & F1 & Accu. \\
		\hline
		OS-AvgSentLen & 0.590 & 0.68 & 0.632 & 0.604 & 0.344 & 0.552 &	0.424 & 0.463 \\
		\hline
		OS-AvgTokenLen & 0.6 & 0.530 & 0.563 & 0.588 & 0.363 & 0.406 &	0.383 & 0.484 \\
		\hline
		OS-Punctuations\_Ratio & 0.541 & 0.756 & 0.630 & 0.557 & 0.249 &	0.587 & 0.35 & 0.354 \\
		\hline
		OF-Unigrams & 0.732 & 0.62 & 0.669 & 0.696 & 0.602 & 0.573 & 0.584 & 0.64 \\
		\hline
		OF-Bigrams & 0.714 & 0.633 & 0.669 & 0.688 & 0.547 & 0.570 & 0.556 & 0.615 \\
		\hline
		OF-Punctuations & 0.650 & 0.638 & 0.643 & 0.646 & 0.468 & 0.558 & 0.507 & 0.555 \\
		\hline
		\method-PFS & 0.804 & 0.646 & 0.716 & 0.744 & 0.707 & 0.616 & 0.658 & 0.705 \\
		\hline
		\method-F & 0.752 & 0.711 & 0.731 & 0.738 & 0.642 & 0.677 & 0.659 & 0.686 \\
		\hline
		\textbf{\method} & 0.778 & 0.726 & \textbf{0.751} & 0.759 & 0.680 & 0.699 & \textbf{0.688} &	0.713 \\
		\hline
	\end{tabular}
	\caption{Performance results on the Amazon dataset.}
	\label{table:Amazon-main-results}
\end{table*}

\begin{table*}[t]
	\captionsetup{skip=4pt}
	\small
	\centering
	\begin{tabular}{l|c|c|c|c||c|c|c|c}
		\hline
		{} & \multicolumn{4}{c||}{\textbf{\textbf{$\textit{eval}_\textit{{cha}}$}}} & \multicolumn{4}{c}{\textbf{$\textit{eval}_\textit{{cp}}$}} \\
		\hline
		Feature Name & Prec. & Recall & F1 & Accu. & Prec. & Recall & F1 & Accu. \\
		\hline
		OS-AvgSentLen & 0.614 & 0.702 & 0.655 & 0.631 & 0.364 & 0.582 &	0.448 & 0.488 \\
		\hline
		OS-AvgTokenLen & 0.647 & 0.538 & 0.587 & 0.622 & 0.343 & 0.381 & 0.361 & 0.496 \\
		\hline
		OS-Punctuations\_Ratio & 0.547 & 0.78 & 0.643 & 0.567 & 0.248 &	0.617 & 0.354 & 0.354 \\
		\hline
		OF-Unigrams & 0.822 & 0.697 & 0.754 & 0.772 & 0.698 & 0.662 & 0.678 & 0.72 \\
		\hline
		OF-Bigrams & 0.790 & 0.698 & 0.741 & 0.756 & 0.648 & 0.655 & 0.651 & 0.693 \\
		\hline
		OF-Punctuations & 0.711 & 0.696 & 0.702 & 0.706 & 0.512 & 0.622 & 0.561 & 0.608 \\
		\hline
		\method-PFS & 0.851 & 0.724 & 0.782 & 0.798 & 0.709 & 0.686 & 0.695 & 0.736 \\
		\hline
		\method-F & 0.825 & 0.781 & 0.8 & 0.805 & 0.684 & 0.748 & 0.711 & 0.738 \\
		\hline
		\textbf{\method} & 0.864 & 0.769 & \textbf{0.813} & 0.824 & 0.745 & 0.742 & \textbf{0.744} & 0.771 \\
		\hline
	\end{tabular}
	\caption{Performance results on the Yelp dataset.}
	\label{table:Yelp-main-results}
\end{table*}

\textbf{Evaluation Schemes and Metrics}: We use two evaluation schemes, \textit{CH-accounts detection} evaluation ($\textit{eval}_\textit{{cha}}$) and \textit{change-point detection } evaluation ($\textit{eval}_\textit{{cp}}$), and report corresponding precision, recall, F1, and accuracy on both tasks. For $\textit{eval}_\textit{{cha}}$, we only identify CH accounts but not the actual change locations. 
For $\textit{eval}_\textit{{cp}}$, we go one step further to also evaluate the identified change locations. Since it is hard to identify the exact change point (the review) at which a change-of-hands has occurred, we define a window $x \pm y$ around the actual change point $x$ with window size $y$, and consider the predicted change point as accurate if it resides within the window. When a change point is detected for a non-CH account, it is considered an error. We study the performance by varying $y$ in Sec.~\ref{sec:result-comparison}.

All evaluation results are based on averaging the results of 5 runs on the constructed datasets. Each time we randomly sample $200$ accounts from each dataset, $100$ in each class, as the development set and use the rest $500$ accounts as the test set. Statistical significance tests are also performed.


\subsection{Baselines}
Since there is no previous work on detecting CH accounts, we propose the following baselines:

\begin{itemize}[leftmargin=*,noitemsep,nolistsep]
	\item \textbf{One Sequence (OS)}. The simplest approach to detecting CH accounts is to construct a single sequence of feature values directly from a moving window of reviews of size $K$ {(one similarity value per review window)} of an account and use it to run a change point detection algorithm. The set of features we tried includes: average sentence length, average token length, ratio of nouns, ratio of adjectives and adverbs, ratio of function words, and ratio of punctuations and special characters. This baseline thus produces 6 results named with OS- as the prefix.
	
	\item \textbf{One Feature (\textit{OF})}. This baseline is a variant of \method. It only uses one of the features from $\boldsymbol{F}^{all}$ as input. Thus, lines 10-11 in Alg.~\ref{algo:1} do not have any effect. Since $\boldsymbol{F}^{all}$ contains 10 features, this baseline produces 10 different results, which are named with OF- as the prefix.
	
	\item \textbf{CHAD~w/out Pivot-level Feature Selection (\textit{CHAD-PFS})}. This baseline is another variant of~\method. It does not perform pivot-level feature selection in Alg.~\ref{algo:1}. It employs the same procedure to pre-select a feature set and thus shares the same $\boldsymbol{F}$ with \method.
	
	\item \textbf{CHAD~w/out Pre-selecting F (\textit{CHAD-F})}. This is a variant of \method~that directly uses $\boldsymbol{F}^{all}$ without pre-selecting feature set $\boldsymbol{F}$ as input.
	
	
\end{itemize}

\noindent
Note that we do not compare with existing spam filtering or sockpuppet detection methods as they all regard reviews from one account as written by a single author. Thus, none of them is able to detect stylistic changes within an account.

\subsection{Parameter Settings}
\label{param-setting}
For each run, we use the respective development set of each dataset to set parameters. For both datasets, $K = 5$ was chosen for the window size in constructing similarity sequences
, and $\lambda_S = 2$ was chosen for vote smoothing. 
For change-point detection, we use the implementation in R \textit{changepoint} package \cite{cpt-journal}, which outputs an estimated change point along with its confidence level. We set a confidence level threshold of $\theta_{conf} = 0.99$ based on the development set, and consider any detected change point with confidence level lower than $\theta_{conf}$ as no change (\textit{none}). In parameter selection and in the main results reporting, we use $y = 5$ for $\textit{eval}_\textit{{cp}}$, and later show the results by varying $y$.

We make the following remarks about these parameters. First, using a very small $K$ (e.g., 1 or 2) leads to bad performance due to the high variance in similarities between reviews. On the other hand, while using a large $K$ (e.g., 7 or 8) improves results for CH accounts detection ($\textit{eval}_\textit{{cha}}$), the performance of change-point detection ($\textit{eval}_\textit{{cp}}$) drops due to loss of granularity. {Second,} it is important to set the confidence level threshold $\theta_{conf}$ high to consider only the most confident detections---due to the fact that each review is unique in some way, and the constructed similarity sequences unavoidably fluctuate to a large extent.


\subsection{Main Results and Analysis}
\label{sec:result-comparison}
We present our main results on Amazon and Yelp datasets, respectively in Tables \ref{table:Amazon-main-results} and \ref{table:Yelp-main-results}. For each dataset we only list the best-3 OS results, best-3 OF results, \method-PFS, \method-F and our proposed \method.

First, \method~significantly outperforms all the baselines ($p < 0.03$) on both datasets. \method-F, which only performs pivot-level feature selection, and \method-PFS, which only performs global feature pre-selection are both worse. 

Second, the best performing OS and OF baselines are consistent on both datasets. The results of OS baselines are quite poor, for which there are two possible explanations. First, they rely on computing a single value as feature, which may not be sufficient in capturing the differences between users in CH accounts. Second, they construct only one sequence, which is less reliable. Although OF baselines generally perform better, they are not reliable for the same reason.

Comparing the results on two datasets, we found that better results are generally achieved on Yelp dataset than on Amazon dataset. 
We believe the difference is mainly caused by the nature of the two datasets. Detecting CH accounts and their change locations are generally harder on Amazon dataset because it contains numerous categories of products. An Amazon reviewer is likely to post reviews on a variety of products, which creates {big} variance when computing similarities. 

\begin{table}[t]
	\captionsetup{skip=4pt}
	\small
	\centering
	\begin{tabular}{l|c|c|c|c}
		\hline
		{$y$} & = 1 & = 3 & = 5 & = 7 \\
		\hline
		\multicolumn{5}{c}{Amazon} \\
		\hline
		\method-PFS & 0.129 & 0.473 & 0.658 & 0.676 \\
		\hline
		\method-F & 0.133 & 0.469 & 0.658 & 0.675 \\
		\hline
		\method & 0.161 & 0.5 & 0.688 & 0.703 \\
		\hline
		\multicolumn{5}{c}{Yelp} \\
		\hline
		\method-PFS & 0.138 & 0.543 & 0.695 & 0.725 \\
		\hline
		\method-F & 0.131 & 0.568 & 0.711 & 0.749 \\
		\hline
		\method & 0.163 & 0.596 & 0.743 & 0.768 \\
		\hline
	\end{tabular}
	\caption{Effect of varying $y$ on $\textit{eval}_\textit{cp}$.}
	\label{table:varying-y}
\end{table}

Lastly, we investigate the effect of window size $y$ (Sec.~\ref{sec:data-and-eval})
under change-point evaluation ($\textit{eval}_\textit{{cp}}$) by varying $y=1,3,5,7$. We report the results for \method-PFS, \method-F, 
and \method~in Table~\ref{table:varying-y}. Only results for $\textit{eval}_\textit{{cp}}$ are listed as those for $\textit{eval}_\textit{{cha}}$ are not affected by $y$. Only average F1 scores are given. As we can see, \method~significantly outperforms the rest methods ($p < 0.03$) regardless of $y$. And as expected, all results improve for increased values of $y$.

\subsection{Experiments in Another Setting}
\label{sec:additional-comparison}
For our main results above, we constructed two datasets in which the CH accounts and non-CH accounts have similar distributions in their sizes (numbers of reviews). In reality, this may not always be the case. CH accounts may in general contain more reviews because spammers may write a lot of fake reviews after purchasing the accounts. In order to test the performance of our method in such cases, we constructed two different datasets without matching the size distributions of CH and non-CH accounts. In particular, for both Amazon and Yelp corpora, we randomly sample 700 original accounts with at least 20 reviews, in which 350 are directly used as non-CH accounts, and the rest are used as the first accounts in CH accounts. Then we randomly sample another 350 accounts with at least 10 reviews as the second accounts in CH accounts. The reason we select at least 20 reviews for the first accounts is because we assume accounts change hands after a sufficiently long history. Statistics of the data is shown in Table~\ref{table:new-dataset-stats}.
We only report F1 scores of two strong baselines (\method-PFS and \method-F), as well as our \method~under both evaluation schemes (Table~\ref{table:new-dataset-result}). As we can see, \method~again performs the best in the new datasets. We also notice the \method~performs better here than in the previous set of experiments (Table~\ref{table:Amazon-main-results} and \ref{table:Yelp-main-results}). 
{The reason is that in the previous experiments setting, there are fewer reviews from the first user/reviewer in the CH accounts, which makes it harder for the algorithm to find reliable patterns.} 

\begin{table}
	\captionsetup{skip=4pt}
	\centering
	\small
	\begin{tabular}{l|c|c|c|c|c}
		\hline
		{} & \textbf{Mean} & \textbf{Med.} & \textbf{Stdev} &\textbf{Min} & \textbf{Max} \\
		\hline
		$Amazon_{NCH}$ & 42.1 & 30 & 30.3 & 20 & 205 \\
		\hline
		$Amazon_{CH}$ & 62.8 & 52 & 34.8 & 30 & 228 \\
		\hline
		${Yelp}_{NCH}$ & 44.4 & 41 & 19.1 & 20 & 85 \\
		\hline
		${Yelp}_{CH}$ & 72.3 & 65 & 24.4 & 30 & 142 \\
		\hline
	\end{tabular}
	\caption{Size statistics of the new data.}
	\label{table:new-dataset-stats}
\end{table}

\begin{table}[t]
	\captionsetup{skip=4pt}
	\small
	\centering
	\begin{tabular}{l|c|c|c|c}
		\hline
		& \multicolumn{2}{c|}{\textbf{Amazon}} & \multicolumn{2}{c}{\textbf{Yelp}} \\
		\hline
		& {\textbf{\textbf{$\textit{eval}_\textit{{cha}}$}}} & {\textbf{$\textit{eval}_\textit{{cp}}$}} & {\textbf{\textbf{$\textit{eval}_\textit{{cha}}$}}} & {\textbf{$\textit{eval}_\textit{{cp}}$}}\\
		\hline
		\method-PFS & 0.805 & 0.726 & 0.844 & 0.752 \\
		\hline
		\method-F & 0.815 & 0.731 & 0.883 & 0.771 \\
		\hline
		\method & 0.816 & 0.753 & 0.886 & 0.786 \\
		\hline
	\end{tabular}
	\caption{Results on the new data ($y=5$).}
	\label{table:new-dataset-result}
\end{table}

\section{Conclusion}
This paper proposed the new problem setting of detecting changed-hands accounts {which complements the existing spammer detection settings and problems. To the best of our knowledge, the problem has not been explored before. The problem presents some unique challenges due to the differences in intra-user and inter-user writing styles. We presented a novel detection algorithm to determine if an account has changed hands and the possible change point.} Extensive experiments on two datasets constructed using Amazon and Yelp review data 
showed that our method outperforms 
a list of baselines significantly.

\bibliographystyle{named}
\bibliography{ijcai18}

\end{document}